\documentclass[conference]{IEEEtran}

\usepackage{graphicx}
\usepackage{amsmath}
\usepackage{algorithm}
\usepackage{algorithmic}

\begin{document}
\title{Can Machine Learn Steganography? \\{\large \emph{- Implementing LSB Substitution and Matrix Coding Steganography with Feed-Forward Neural Networks}}}
\author{\IEEEauthorblockN{Han-Zhou Wu, Hong-Xia Wang and Yun-Qing Shi}
\IEEEauthorblockA{Contact email: h.wu.phd@ieee.org}
\IEEEauthorblockA{* This manuscript may be revised and updated in the future.}}

\maketitle
\begin{abstract}
In recent years, due to the powerful abilities to deal with highly complex tasks, the artificial neural networks (ANNs) have been studied in the hope of achieving human-like performance in many applications. Since the ANNs have the ability to approximate complex functions from observations, it is straightforward to consider the ANNs for steganography. In this paper, we aim to implement the well-known LSB substitution and matrix coding steganography with the feed-forward neural networks (FNNs). Our experimental results have shown that, the used FNNs can achieve the data embedding operation of the LSB substitution and matrix coding steganography. For steganography with the ANNs, though there may be some challenges to us, it would be very promising and valuable to pay attention to the ANNs for steganography, which may be a new direction for steganography. 
\end{abstract}


\IEEEpeerreviewmaketitle
\section{Motivation}
In machine learning, artificial neural networks (ANNs) are a family of computational models inspired from neuroscience, which can be used to
approximate functions that may depend on a huge number of inputs and are usually unknown. As efficient models for statistical pattern recognition, 
ANNs have been widely applied in applications. A most common model is the feed-forward neural network (FNN) [1], in which 
the information moves along one direction, forward, from the input neurons, through the hidden neurons (if any) and to the output neurons. Moreover, there is no cycle or loop in a FFN, which is different from another kind of neural network called recurrent neural network (RNN).

Due to the powerful abilities to deal with highly complex and ill-defined problems, ANNs have been studied for many years in the hope of achieving human-like performance in lots of fields such as regression analysis, object classification, data/signal processing, robotics and control. For example, in recent years, deep ANNs [2] have won numerous competitions in the field of pattern recognition and machine learning. The deep ANNs have greatly advanced the application fields such as visual recognition systems, medical diagnosis, financial applications, data mining, e-mail spam filtering, game-playing and decision making. Just recently, with deep ANNs, a Google computer program named \emph{AlphaGo} [3] even beat a professional human player at the 2-player board game of \emph{Go}, which has around $10^{170}$ possible positions.

Steganography [4] is referred to the art of embedding secret data into innocent objects such as image, video and audio. It is required that, there should be impossible for an eavesdropper to distinguish ordinary objects and objects containing secret data. Unlike cryptography, since steganography even conceals the presence of covert communication, it has been widely used to provide secret communication. Steganographic algorithms for images usually embed secret data into a cover image by altering the pixel values without introducing obvious artifacts. As the resultant image called \emph{stego image} hides the existence of secret data, 
only the recipient who owns the secret key can perfectly retrieve the embedded information.
\begin{figure}[!t]
\centering
\includegraphics[width=3.5in]{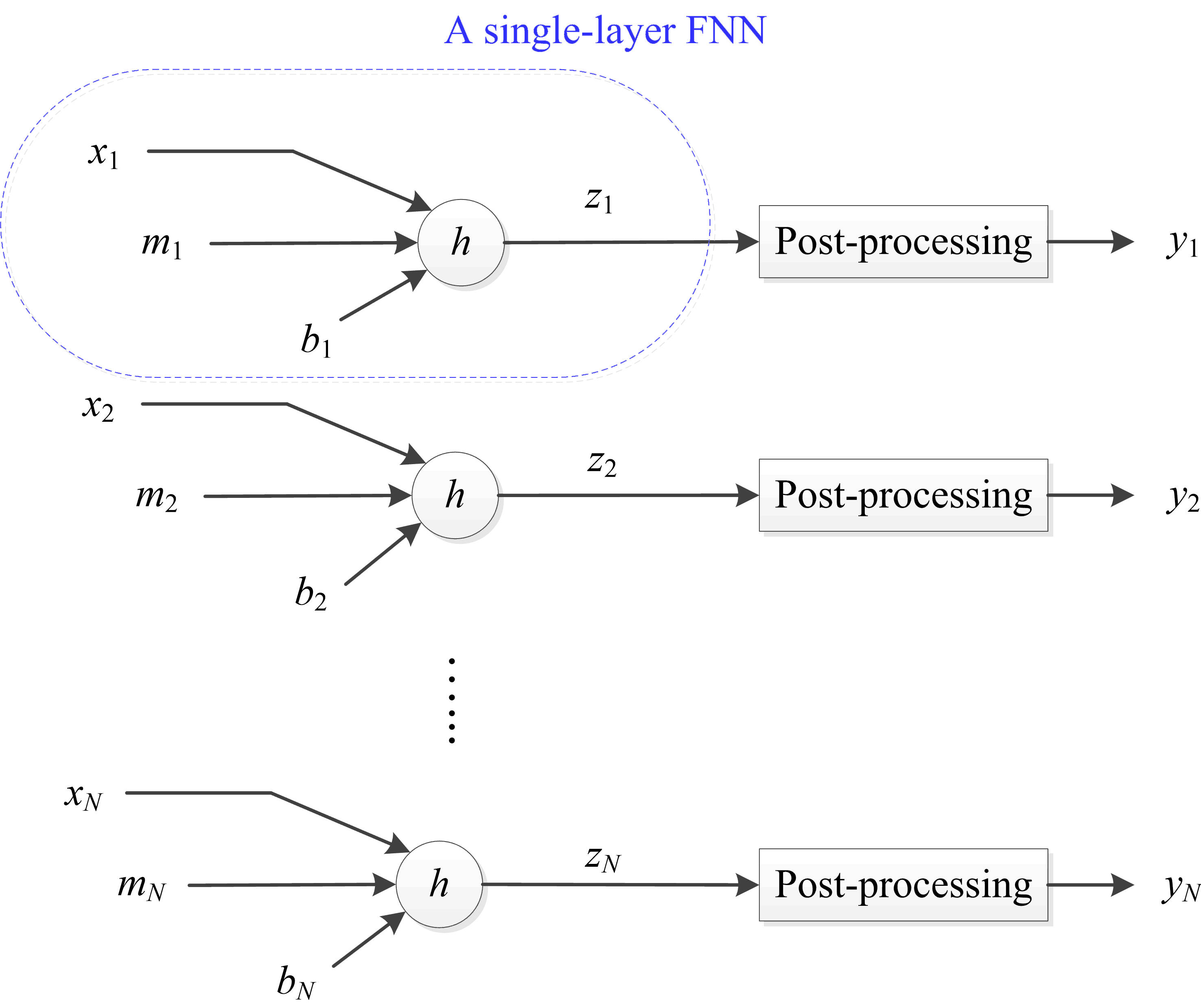}
\caption{Implement the LSB substitution with single-layer FNNs: $n_1 = 1$.}
\end{figure}
For a steganographic system, the stego image can be considered as the output of an encoding function that relies on the cover image, the secret data and a secret key. The decoding function aims to reconstruct the secret data according to the stego image and secret key. Since the ANNs have the ability to approximate functions from observations, it is straightforward to consider the ANNs for steganography. And, it is very likely to apply the ANNs for the data embedding or data extraction process of a steganographic system. For example, a steganographic algorithm may adopt the ANNs to find the optimal data embedding strategy (e.g., adaptively select the most suitable pixels to be embedded), while subjected to an upper-bounded distortion. Also, since the ANNs has the ability to approximate very complex functions, any steganographic attacker should not be able to generate a legal stego image, extract the hidden information, or even hardly locate the probably embedded image regions. Furthermore, in future, there may be no need for a human to design the detailed data embedding and data extraction functions, as they may be done by training the ANNs to achieve these features. While the above scenarios may seem to be incredible or even impossible, it would be very promising and valuable to pay attention to the ANNs for steganography. 

In this paper, we aim to implement the LSB substitution and matrix coding steganography with the FNNs. Though both can be easily done  by applying the FNNs, we can draw some valuable arguments from the experiments, which leads us to study ANN-based steganography in a good direction. The rest of this paper are organized as follows. Section II shows the details to achieve the LSB substitution with the FNNs, followed by the matrix coding steganography in Section III. We present some discussion and analysis in Section IV. Finally, we conclude this paper in Section V. 
\section{Implementing LSB Substitution with FNNs}
\subsection{LSB Substitution Steganography}
Let $\textbf{x}=(x_1, x_2, ..., x_N)$ and $\textbf{y}=(y_1, y_2, ..., y_N)$ denote the cover vector and stego vector. The LSB substitution hides the secret data by simply replacing the LSB of each $x_k$ with the secret bit to be embedded. Hence, the data embedding function can be described as
\begin{equation}
y_k=f(x_k,m_k)=x_k-\textrm{LSB}(x_k)+m_k, (1\leq k\leq N).
\end{equation}
Here, $m_k\in\{0, 1\}$ denotes the \emph{k}-th secret bit to be embedded.

When an intended receiver acquires $\textbf{y}$, he is able to retrieve the hidden information by collecting the stego LSBs, i.e.,
\begin{equation}
m_k=f^{-1}(y_k)=\textrm{LSB}(y_k), (1\leq k\leq N).
\end{equation}

Note that, though theoretically $x_k$ can be arbitrary integer, only the LSB of $x_k$ works for data hiding. It indicates that, instead of $x_k$, we can consider the LSBs of \textbf{x} as the input. For simplicity, we will consider both \textbf{x} and \textbf{y} as a binary vector.
\subsection{Using Single-Layer FNNs} 
A single-layer FNN includes one input layer and one output layer of processing units. There is no hidden layers between the input layer and output layer. Let $\textbf{I}^{(k)} =({a_1}^{(k)}, {a_2}^{(k)}, ..., {a_{n_k}}^{(k)})$ and $\textbf{W}^{(k)} = \{{w_{i,j}}^{(k)}|1\leq i \leq n_{k-1},1\leq j\leq n_{k}\}$ denote the output vector of the $k$-th layer and the input weights for the $k$-th layer. For example, for a single-layer FNN, $\textbf{I}^{(0)}$ and $\textbf{W}^{(1)}$ represent the output vector of the input layer (or called the input vector of the output layer), and the input weights for the output layer. For the $k$-th layer in a fully-connected FNN, we have
\begin{equation}
{a_j}^{(k)} = {h_j}^{(k)}(\sum_{i=1}^{n_{k-1}} w_{i,j}\cdot {a_i}^{(k-1)}),(1\leq j< n_k).
\end{equation}
where ${h_j}^{(k)}(\cdot)$ represents the activation function. Note that, in default, we consider ${a_{n_k}}^{(k)} = 1$ as the bias input.
\begin{figure}[!t]
\centering
\includegraphics[width=3.5in]{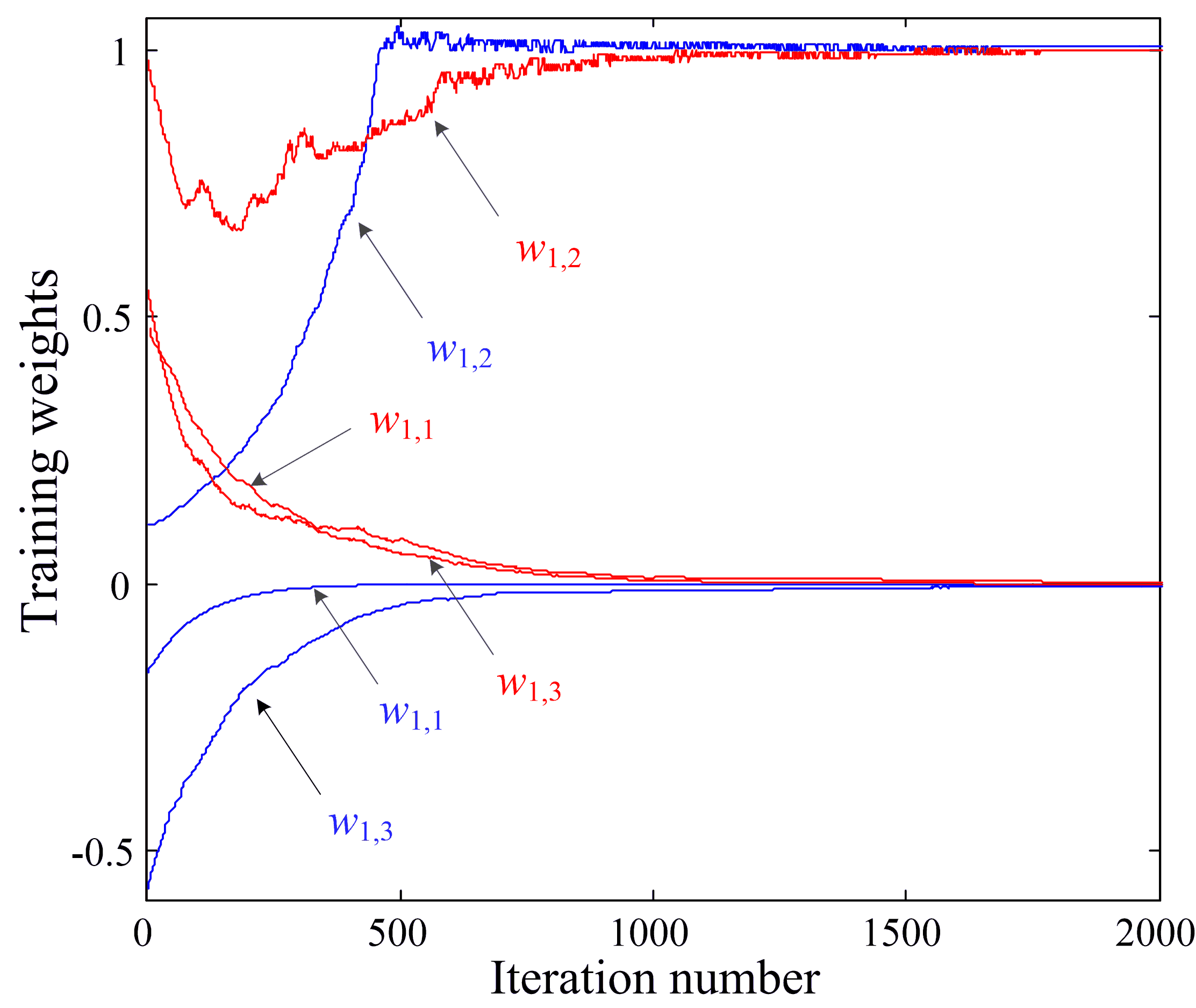}
\caption{The training weights versus iteration number for two sets of randomly initialized weights ($\delta$ = 0.01).}
\end{figure}
We present two methods to implement the LSB substitution with single-layer FNNs. One is to set $n_1 = 1$, and $n_1 > 1$ for the other one. 
We use $h(x) = x$ as the activation function in default. Note that, we can also use other activation functions.
In case $n_1 = 1$, we process each of the cover elements \{$x_1, x_2, ..., x_N$\} with the identical single-layer FNN to produce the \emph{N} stego elements \{$y_1, y_2, ..., y_N$\}. Fig. 1 shows the sketch for the case $n_1 = 1$. In Fig. 1, $b_i = 1~(1\leq i\leq N)$ denote the bias inputs. 
For each triple-input ($x_i,m_i,b_i$), ($1\leq i\leq N$), we aim to find out reliable triple-weights ($w_{i,1}, w_{i,2}, w_{i,3}$) such that $y_i$ can be correctly determined. All the triple-inputs will have the identical triple-weights, meaning that, we should only compute 
($w_{1,1}, w_{1,2}, w_{1,3}$). We are to compute the weights by training observations. For consistency, we think of the used observations as the training data, and the testing data consist of \textbf{x} and \textbf{m} = $(m_1, m_2, ..., m_N)$. We employ randomly generated observations to quickly find the weights. Specifically, we randomly generate a certain number of bit-pairs (${{x}'}_i, {{m}'}_i$), ($1\leq i\leq T$), and their target outputs ${{z}'}_i$, ($1\leq i\leq T$). For each $\{{{x}'}_i, {{m}'}_i, {{z}'}_i\} \in \{0, 1\}^3$, it should meet ${{z}'}_i = {{m}'}_i$ as required by the LSB substitution operation. The randomly generated training data are individually fed into the single-layer FNN to train the required parameters. Due to its simplicity, during the training, we use the \emph{hill climbing} [5] optimization technique to optimize the parameters, which can be described as Algorithm 1.

Fig. 2 shows the training weights versus iteration number by applying the Algorithm 1 for two sets of randomly initialized weights, where ($w_{1,1}, w_{1,2}, w_{1,3}$) converge to (0, 1, 0). The optimized weights can be used for data hiding. For example, with randomly initialized weights (0.482945, 0.979194, 0.550665), we have the optimized weights (0.000410723, 0.999325, 2.03115$\times 10^{-5}$) by applying 6000 iterations. For any $x_i = 1$ and $m_i = 1$, we have $z_i = 0.9998$. By setting a threshold such as $\theta = 0.5$, we can obtain the stego value $y_i = 1$ as $z_i \geq \theta$.  

In case $n_1 > 1$, we divide both \textbf{x} and \textbf{m} into $n/n_1$ disjoint subvectors (suppose that $n$ is divisible by $n_1$). Fig. 3 shows the sketch to implement the LSB substitution with single-layer FNNs in case $n_1 > 1$. Similarly, we use randomly generated training data and employ the gradient descent [6] technique to optimize the weights. Our experimental results have shown the FNN architecture in Fig. 3 can be successfully applied for data hiding. For example, \emph{Appendix A} provides the Python code to optimize the weights in case $n_1 = 3$.
\\
---------------------------------------------------------------------------\\
Algorithm 1: Optimizing weights with hill climbing technique.\\
1.~~~Initialize \textbf{w} = ($w_{1,1}, w_{1,2}, w_{1,3}$) $\in [-1, 1]^3$ and $\delta\in (0, 1)$\\
2.~~~\textbf{For} $i = 1,2,...,T$ \textbf{do}\\
3.~~~~~~Randomly set $({{\textbf{v}}'}_i; {{z}'}_i) = ({{x}'}_i, {{m}'}_i, 1; {{m}'}_i) \in \{0, 1\}^4$\\
4.~~~~~~Set ${\textbf{w}}' = \textbf{w}$\\
5.~~~~~~\textbf{For} all possible \textbf{e} = ($e_1, e_2, e_3$) $\in \{-1, 0, 1\}^3$ \textbf{do}\\
6.~~~~~~~~~\textbf{If} $Loss({{\textbf{v}}'}_i, {{z}'}_i; \textbf{w}+\delta\cdot\textbf{e}) < Loss({{\textbf{v}}'}_i, {{z}'}_i; {\textbf{w}}')$ \textbf{do}\\
7.~~~~~~~~~~~~Set ${\textbf{w}}' = \textbf{w}+\delta\cdot\textbf{e}$\\
8.~~~~~~Set \textbf{w} = ${\textbf{w}}'$\\
9.~~~\textbf{Return} \textbf{w}\\
---------------------------------------------------------------------------
\subsection{Using Multi-Layer FNNs}
A multi-layer FNN has one input layer, one output layer, and one or more hidden layers of processing units. 
The LSB substitution can be also implemented by using multi-layer FNNs (in fact, the single-layer FNN can be considered as the special case of a multi-layer FNN). Fig. 4 shows a multi-layer FNN designed for LSB substitution. The designed FNN can be successfully used for data hiding. For example, \emph{Appendix B} provides the Python code to optimize the weights for the multi-layer FNN with one hidden layer. It is noted that, when applied to gray-scaled images, we can use the whole pixel LSBs to be embedded as a part of the input of a multi-layer FNN. Also, all the pixel LSBs to be embedded can be divided into disjoint pixel vectors and then separately fed into the identical multi-layer FNN. We here prefer to recommend the latter since the training time will be reduced, comparing with the former.

\section{Implementing Matrix Coding with FNNs}
The well-known matrix coding used in [7] is an application of the binary Hamming code. It can embed \emph{k} bits into $2^k-1$ cover pixels by changing only one LSB with the probability of $(2^k-1)/2^k$, which results in an average distortion $1/2^k$ with an embedding rate $k/(2^k-1)$ and embedding efficiency $k\cdot 2^k/(2^k-1)$. Let $\textbf{x}=(x_1, x_2, ..., x_{2^k-1})$ and $\textbf{m}=(m_1, m_2, ..., m_k)$ denote the cover vector and the bit-vector to be embedded. In matrix coding, the hash function is defined as
\begin{equation}
f(\textbf{x}) = \bigoplus _{i=1}^{2^k-1}x_i\cdot i.
\end{equation}
The bit place is then computed as
\begin{equation}
s = f(\textbf{x})\bigoplus \textbf{x}.
\end{equation}
that we have to change, which results in the stego vector
\begin{equation}
\textbf{y}=\left\{\begin{matrix}
\textbf{x},~\text{if}~s=0~(\rightleftharpoons \textbf{x}=f(\textbf{x}));\\ 
(x_1,x_2,...,1-x_s,...,x_{2^k-1}),~\text{otherwise}.
\end{matrix}\right.
\end{equation}

We implement the matrix coding operation with a multi-layer FNN. Fig. 5 shows the used FNN structure. It is noted that, the used activation functions in the hidden layers will be the sigmoid function (other non-linear functions may be also usable) since a completely linear FNN can not achieve \emph{the XOR operation}. Our experiments have shown that the used FNN structure provides correct outputs, e.g., \emph{Appendix C} provides the Python code to optimize the weights for the multi-layer FNN with one hidden layer in case \emph{k} = 2. Similarly, it is recommended that, when applied to images, we should use multiple disjoint cover vectors for finally generating the stego images since it is indeed time consuming, when to collect the entire cover LSBs as the input of the FNN. That is, in terms of training time with gradient descent, the saturating nonlinearity is very slow [8].
\section{Discussion and Analysis}
In Section II and III, we present the FNN structures used for both the LSB substitution and matrix coding. Though both can be easily done 
by applying the FNNs, we can draw some valuable arguments in this section.
\begin{figure}[!t]
\centering
\includegraphics[width=3.5in]{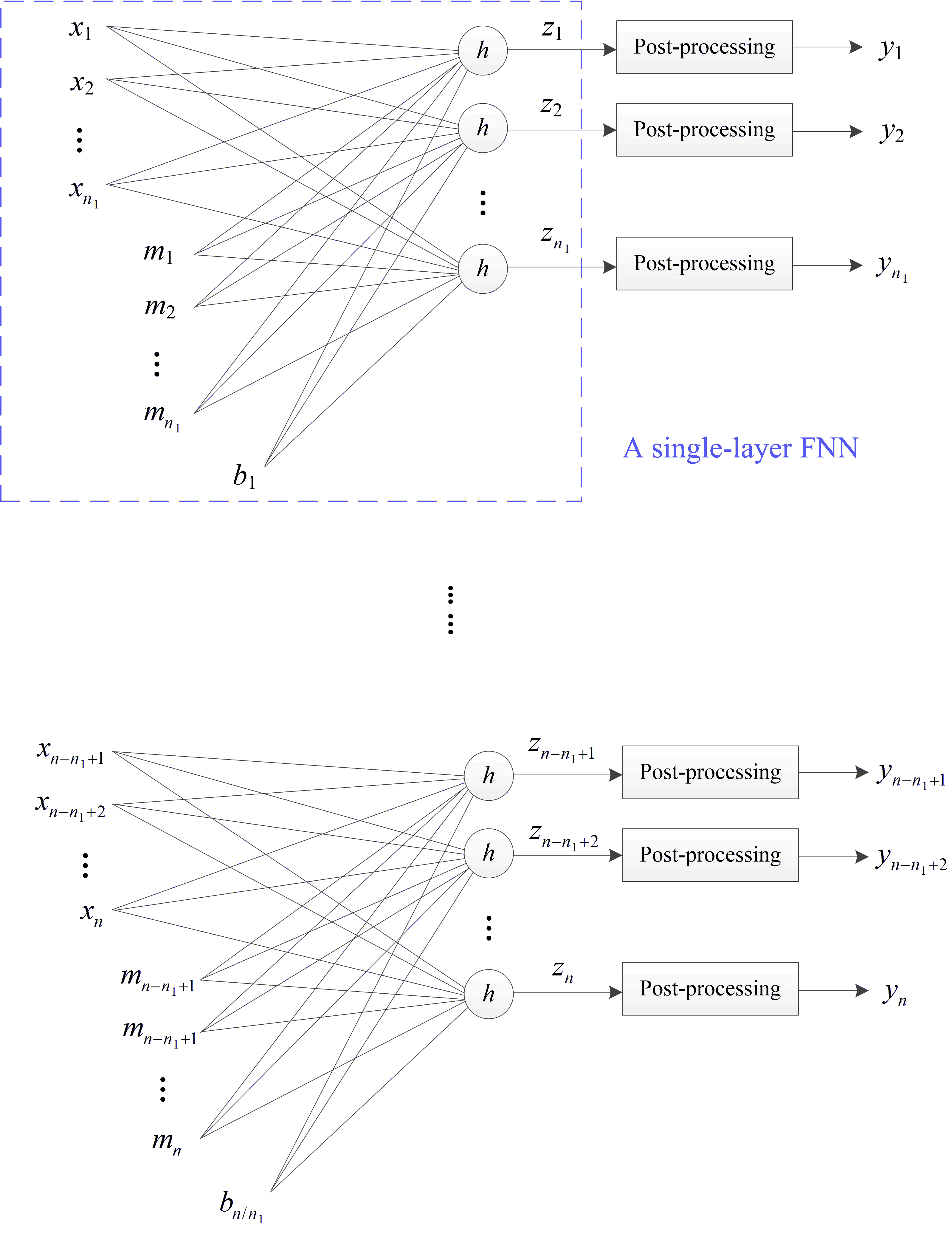}
\caption{Implement the LSB substitution with single-layer FNNs: $n_1 > 1$.}
\end{figure}
\begin{figure}[!t]
\centering
\includegraphics[width=3.5in]{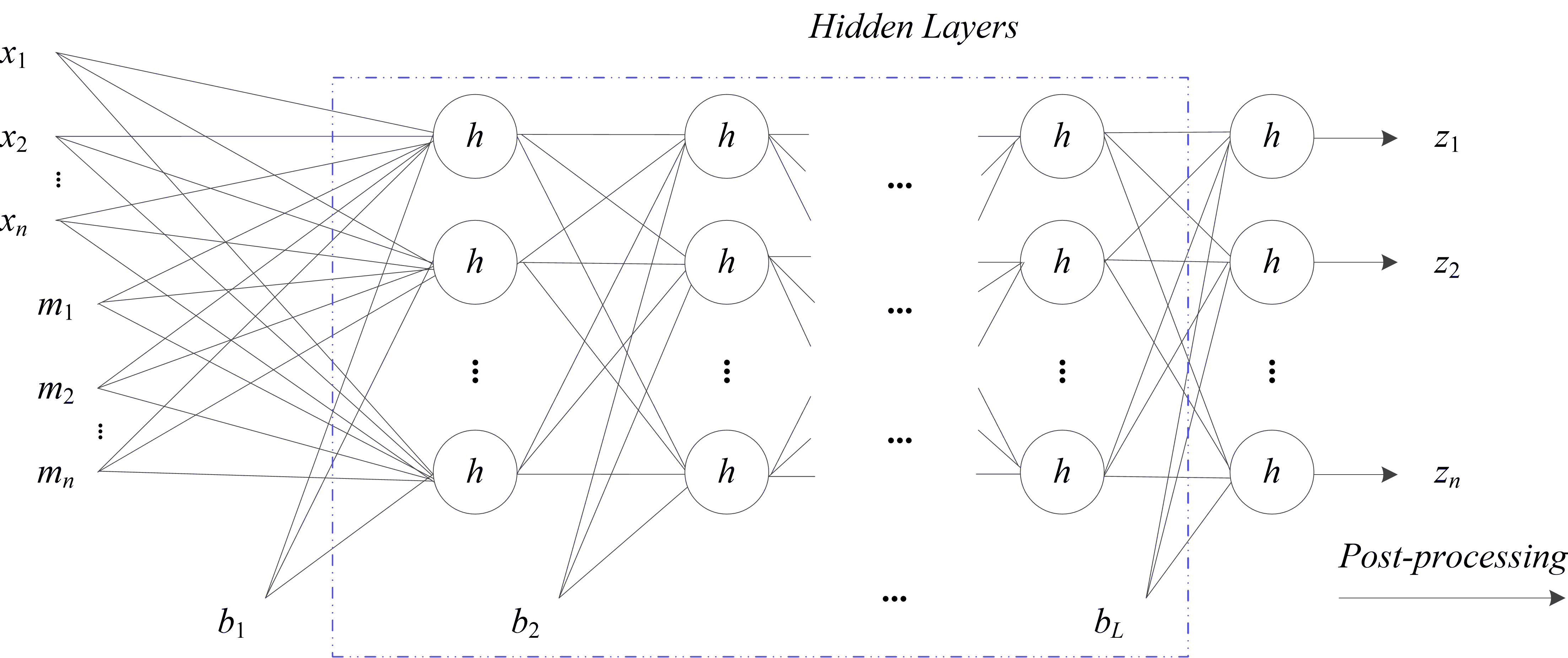}
\caption{Implement the LSB substitution with a fully-connected multi-layer FNN. The number of hidden layers and hidden nodes are indefinite.}
\end{figure} 
\begin{figure}[!t]
\centering
\includegraphics[width=3in]{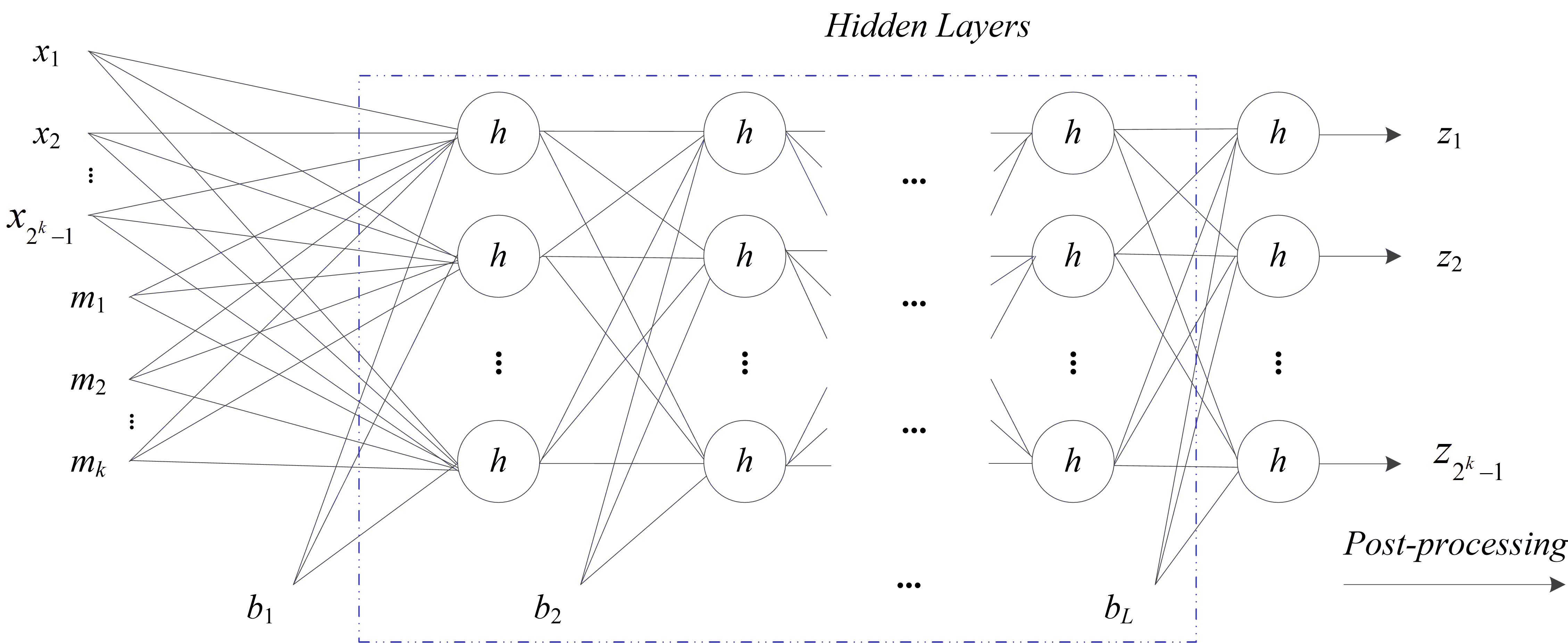}
\caption{Implement the matrix coding with a fully-connected multi-layer FNN. The number of hidden layers and hidden nodes are indefinite.}
\end{figure}

At first, if the data embedding process of a steganographic system can be modelled by an ANN, the usable ANN should be not unique. The reason is that, we can always modify/extend the ANN (e.g., changing the activation functions) or insert never-activated perceptrons into an ANN such that the new ANN still works well for data hiding. Therefore, there usually exists a lot of freedom to design a usable neural network for steganography. Secondly, on the receiver side, we can still employ the ANNs to extract the hidden information. For example, for the LSB substitution, the designed FNN for a decoder should be trained by using the stego vector as input and message vector as output.

Thirdly, in our designed FNNs, we did not take into account the secret key. The reason is that, the secret data \textbf{m} should be encrypted according to the secret key, which can maintain the security of the secret message. On the other hand, though the secret key can help us to find the suitable regions of data-embedding, e.g., a secret key can control the process to (adaptively) find optimal data-embedding positions, we are expecting to achieve this feature by designing an independent neural network in the future. That is, we expect to merge two neural networks for steganography. One is called \textbf{Policy Network (PN)}, and the other is \textbf{Embedding Network (EN)}. As shown in Fig. 6, the PN will be used for finding the optimal data-embedding regions from a cover image, and the EN for data embedding in selected pixels. We are moving our research ahead along this direction. 

In addition, when an ANN-based steganography uses neural networks for data embedding or data extraction, there may exist the problem of bit-embedding error rate (BembER) or bit-extraction error rate (BextER). It indicates that, we may need to adjust the designed ANN structure or employ some robust technologies to correct/reduce the errors. 
\begin{figure}[!t]
\centering
\includegraphics[width=3.5in]{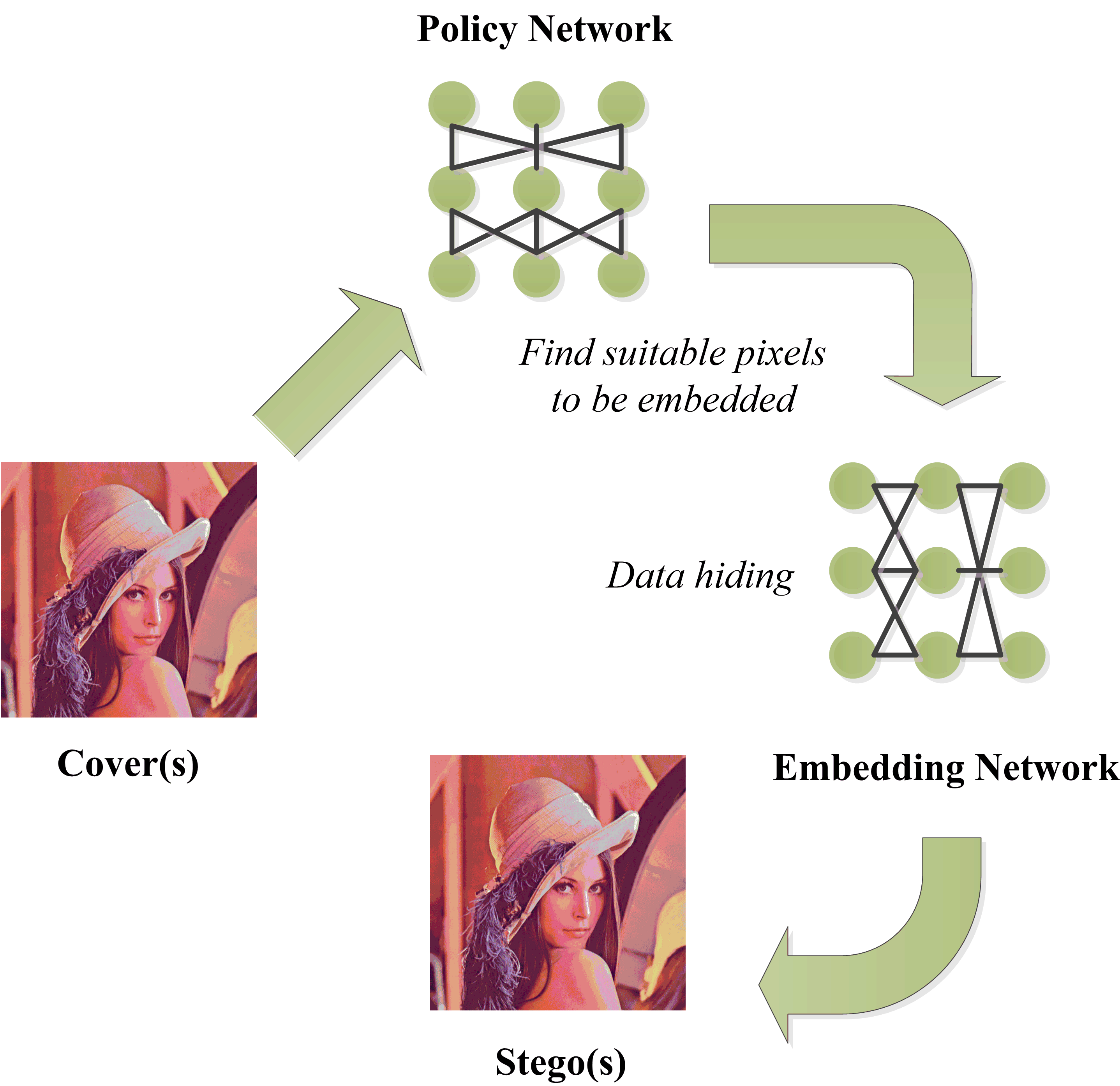}
\caption{Steganographic framework by utilizing a policy network (PN) and an embedding network (EN).}
\end{figure}

\section{Conclusion}
Steganalysis [9], from an opponent's perspective, aims to detect the presence of covert communication caused by steganography. 
Recently, steganalysis by exploiting deep convolutional neural network (CNN) [10, 11] has been studied and achieved competitive detection performance, 
It indicates that steganalysis with deep neural networks has the potential to provide the best detection performance in the future. 
Since the ANNs have the ability to approximate very complex functions from observations, it is possible to use the ANNs for steganography. In this paper, we present some preliminary studies on steganography based on the ANNs. Though there may be some challenges to us, it is desirable to pay attention to ANNs for steganography. Once steganography with neural networks has been deeply studied and achieved excellent results, the war between steganography and steganalysis may be the war among neural networks.

\section*{Acknowledgement}
This work was partly supported by the Chinese Scholarship Council (CSC) under the grant No. 201407000030.

\appendices
\section{Python code for LSB substitution with single-layer FNNs in case $n_1 = 3$}
from pybrain.structure import * \# lib: pybrain.org

fnn = FeedForwardNetwork();

inLayer = LinearLayer(7, name = 'inLayer');

outLayer = LinearLayer(3, name = 'outLayer');

fnn.addInputModule(inLayer);

fnn.addOutputModule(outLayer);

in\_to\_out = FullConnection(inLayer, outLayer);

fnn.addConnection(in\_to\_out);

fnn.sortModules();

from pybrain.supervised.trainers import BackpropTrainer

from pybrain.datasets import SupervisedDataSet

ds = SupervisedDataSet(7, 3);

from random import randint

for ind in range(0, 8000):

$~~~~~$x = [randint(0, 1), randint(0, 1), randint(0, 1)];
	
$~~~~~$m = [randint(0, 1), randint(0, 1), randint(0, 1)];
	
$~~~~~$b = 1;
	
$~~~~~$z = m;
	
$~~~~~$ds.addSample([x[0],x[1],x[2],m[0],m[1],m[2],b], z);

trainer = BackpropTrainer(fnn, ds, verbose = False, learningrate = 0.01);

trainer.trainUntilConvergence(maxEpochs = 7000);

for ind in range(0, 32): \# \emph{user test}
	
$~~~~~$x = [randint(0, 1), randint(0, 1), randint(0, 1)];
	
$~~~~~$m = [randint(0, 1), randint(0, 1), randint(0, 1)];
	
$~~~~~$b = 1;
	
$~~~~~$z = m;
	
$~~~~~$print(z,fnn.activate([x[0],x[1],x[2],m[0],m[1],m[2],b]));

\section{Python code for LSB substitution with a multi-layer FNN containing one hidden layer}
from pybrain.supervised.trainers import BackpropTrainer

from pybrain.tools.shortcuts import buildNetwork

from pybrain.structure.modules import LinearLayer

net = buildNetwork(6, 5, 3, bias = True,  hiddenclass = LinearLayer);

from pybrain.datasets import SupervisedDataSet

ds = SupervisedDataSet(6, 3);

from random import randint

for ind in range(0, 8000):

$~~~~~$x = [randint(0, 1), randint(0, 1), randint(0, 1)];
	
$~~~~~$m = [randint(0, 1), randint(0, 1), randint(0, 1)];
	
$~~~~~$z = m;
	
$~~~~~$ds.addSample([x[0],x[1],x[2],m[0],m[1],m[2]], z);

trainer = BackpropTrainer(net, ds, verbose = False, learningrate = 0.01);

trainer.trainUntilConvergence(maxEpochs = 7000);

for ind in range(0, 32): \# \emph{user test}
	
$~~~~~$x = [randint(0, 1), randint(0, 1), randint(0, 1)];
	
$~~~~~$m = [randint(0, 1), randint(0, 1), randint(0, 1)];
	
$~~~~~$z = m;

$~~~~~$print (z, net.activate([x[0],x[1],x[2],m[0],m[1],m[2]]));

\section{Python code for matrix coding with a multi-layer FNN using sigmoid activation function}
from pybrain.supervised.trainers import BackpropTrainer

from pybrain.tools.shortcuts import buildNetwork

from pybrain.structure.modules import SigmoidLayer, LSTMLayer, LinearLayer

net = buildNetwork(5, 12, 3, bias = True,  hiddenclass = SigmoidLayer);

from pybrain.datasets import SupervisedDataSet

ds = SupervisedDataSet(5, 3);

from random import randint

for ind in range(0, 2000):

$~~~~~$x = [randint(0, 1), randint(0, 1), randint(0, 1)];

$~~~~~$m = [randint(0, 1), randint(0, 1)];

$~~~~~$z = x;

$~~~~~$if (z[0]+z[1])\%2!=m[0] and (z[0]+z[2])\%2==m[1]:

$~~~~~~~~$z[1] = 1 - z[1];

$~~~~~$if (z[0]+z[1])\%2!=m[0] and (z[0]+z[2])\%2!=m[1]:

$~~~~~~~~$z[0] = 1 - z[0];

$~~~~~$if (z[0]+z[1])\%2==m[0] and (z[0]+z[2])\%2!=m[1]:

$~~~~~~~~$z[2] = 1 - z[2];

$~~~~~$ds.addSample([x[0], x[1], x[2], m[0], m[1]], z);

trainer = BackpropTrainer(net, ds, verbose = False, learningrate = 0.01);

trainer.trainUntilConvergence(maxEpochs = 1000);

for ind in range(0, 128): \# \emph{user test}

$~~~~~$x = [randint(0, 1), randint(0, 1), randint(0, 1)];

$~~~~~$m = [randint(0, 1), randint(0, 1)];

$~~~~~$z = x;

$~~~~~$if (z[0]+z[1])\%2!=m[0] and (z[0]+z[2])\%2==m[1]:

$~~~~~~~~$z[1] = 1 - z[1];

$~~~~~$if (z[0]+z[1])\%2!=m[0] and (z[0]+z[2])\%2!=m[1]:

$~~~~~~~~$z[0] = 1 - z[0];

$~~~~~$if (z[0]+z[1])\%2==m[0] and (z[0]+z[2])\%2!=m[1]:

$~~~~~~~~$z[2] = 1 - z[2];

$~~~~~$print (x,m,z,net.activate([x[0],x[1],x[2],m[0],m[1]]));

% that's all folks
\end{document}